\documentclass[fleqn,12pt,twoside]{article}
\usepackage{espcrc1}

\usepackage{graphicx}
\usepackage[figuresright]{rotating}

\newcommand{\AmS}{{\protect\the\textfont2
  A\kern-.1667em\lower.5ex\hbox{M}\kern-.125emS}}

\title{The Experimental Search for Pentaquark}

\author{P. Rossi \address {Laboratori Nazionali di Frascati - INFN, \\ 
        Via Enrico Fermi 40, 00044 Frascati, Italy}
\\for the Jefferson Lab CLAS Collaboration }
       
\begin{document}

\maketitle

\begin{abstract}
The existence of an anti-decuplet of pentaquark particles has been predicted some year ago within the chiral soliton model.
In the last year, several experimental groups have reported evidence for a S=+1 baryon resonance, with mass ranging from 1.52 and 1.55 GeV and width less than 25 MeV, by looking at the invariant mass of the $K N$ system. This resonance, has been identified with the lowest mass of the anti-decuplet, the $\Theta^+$. At the same time, there are a number of experiment, mostly at high energies, that report null results.\\
An overview of the experimental results so far obtained will be given here together with a review of the second generation experiments currently ongoing and planned at Jefferson Lab Hall B.
\end{abstract}

\section{INTRODUCTION}
\begin{figure}[h]
\begin{minipage}[t]{70mm}
\mbox{}\\
\includegraphics[width=7.3cm]{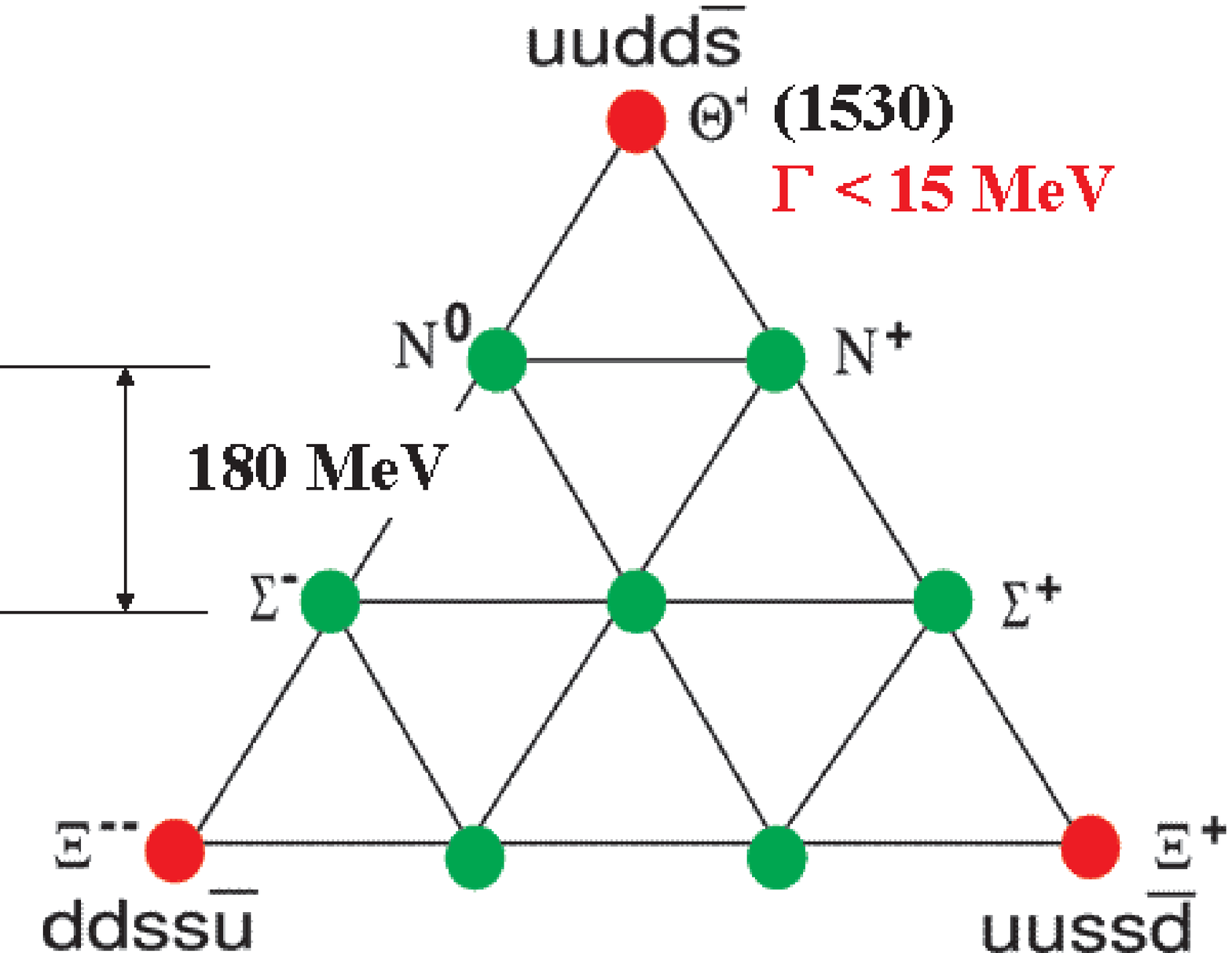}
\vspace{0.3cm}
\caption{The anti-decuplet of pentaquark states as predicted in the Chiral Soliton Model \cite{Diak}.}
\label{fig:decupletto}
\end{minipage}
\hspace{\fill}
\begin{minipage}[t]{70mm}
\mbox{}\\
\includegraphics[width=7.3cm]{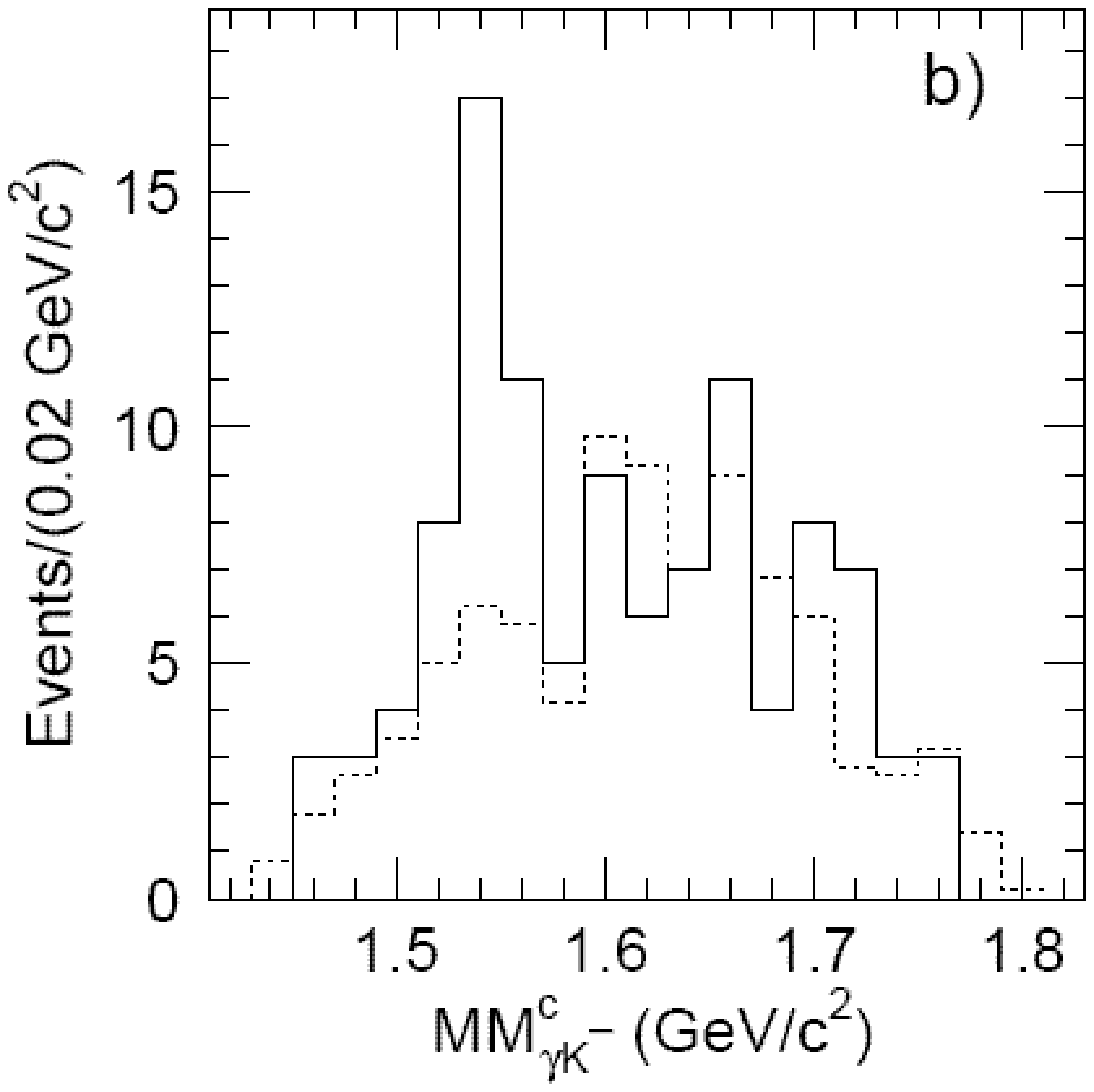}
\vspace{-1.5cm}
\caption{$K^+ n$ invariant mass measured by the LEPS Collaboration \cite{leps} in $\gamma^{12}C \rightarrow K^+ K^- X$ events.}

\label{fig:leps}
\end{minipage}
\end{figure}

All the well-established particles can be categorized using the constituent quark model which describes light mesons as bound states of $q \bar{q}$ pairs, and baryons as bound 3-quarks states.
On the other hand, high energy experiments have shown a more complicated internal structure of mesons and baryons made of a swarms of quarks, anti-quarks and gluons. It is then natural to ask wether particles with more complex configurations exists, like for example 5-quarks ($q q q q \bar{q}$) states, where the $\bar{q}$ has different flavor than the others quarks. These states, with quark content other than $q \bar{q}$ or $q q q $ are termed as {\it exotics}.\\
The idea of exotics has in fact been proposed since the early 70's but the experimental signals for exotic baryons were so controversial that never rised to a level of certainty sufficient for the Particle Data Group's tables \cite{Roos,Jennings}. Till, in its 1988 review the Particle Data Group officially put the subject to sleep \cite{Yost}.\\
Although the lack of clear evidence of exotic particles, theoretical work on this subject was continued by several authors on the basis of quark and bag models \cite{Jaffe} and on the Skyrme model \cite{Manohar,Chemtob}.
Using the latter one, Praszalowicz \cite{Pras} provided the first estimate of the mass of the lightest exotic state, $ M \sim 1530$ MeV, and in 1997 Diakonov, Petrov and Polyakov \cite{Diak}, in the framework of the Chiral Quark Soliton Model, predicted an antidecuplet of 5-quarks baryons, with spin and parity $J^{\pi}=1/2^+$ illustrated in fig.~\ref{fig:decupletto}. The lowest mass member is an isosinglet state, dubbed $\Theta^+$, with quark configuration ($u u d d \bar{s}$) giving S=+1, with mass $\sim 1.54$~GeV and width of around 15 MeV.

Experimental evidence for a S=+1 baryon resonance with mass 1.54 GeV and width less than 25 MeV has been reported for the first time by the LEPS Collaboration at SPring-8 \cite{leps} in the photoproduction on neutron bound in a carbon target.
Immedialely after, several other experimental groups analyzing previously obtained data, have found this exotic baryon in both his decaying channels $\Theta^+ \rightarrow p K^0$ and $n K^+$ \cite{diana,ks,stepan,kouba,saphir,hermes,neutrino,svd,cosy,zeus}.

\section{CURRENT EXPERIMENTAL STATUS}
The properties of the observed candidate pentaquark signals obtained studying different reactions with different experimental methods, are summarized in Table~\ref{table:1}.
\begin{table}[htb]
\caption{Summary table of the experimental results of the different $\Theta^+$ experiments (first column). The $\Theta^+$ decay channels studied are reported in the second column; mass, width and statistical significance of the measured signals in columns 3 to 5.}
\label{table:1}
\newcommand{\m}{\hphantom{$-$}}
\newcommand{\cc}[1]{\multicolumn{1}{c}{#1}}
\renewcommand{\tabcolsep}{1.1pc} 
\renewcommand{\arraystretch}{1.2} 
\begin{tabular}{@{}lllll}
\hline
EXP. & REACTION           & \cc{M (MeV)} & \cc{$\Gamma (MeV)$} & \cc{STAT.} \\
     & CHANNEL           &                                     &                     &     SIG.\\
\hline
LEPS \cite{leps}  &$\gamma n (C^{12}) \rightarrow \Theta^+ K^-$ &  \m1540 $\pm10$ & $<25$ & \m4.6$\sigma$ \\
DIANA \cite{diana} &$K^+ Xe \rightarrow \Theta^+ X$ & \m1539 $\pm2$ & $<9$ & \m4.4$\sigma$ \\
CLAS(d) \cite{stepan} &$\gamma d  \rightarrow \Theta^+ K^- p$ &  \m1542 $\pm 5$ & $<21$ & \m5.2$\sigma$ \\
CLAS(p) \cite{kouba} &$\gamma p  \rightarrow \Theta^+ K^- \pi^+$ &  \m1555 $\pm 1 \pm 10$ & $<26$ & \m7.8$\sigma$ \\
SAPHIR \cite{saphir} &$\gamma p  \rightarrow \Theta^+ K^0 $ &  \m1540 $\pm 4 \pm 2$ & $<25$ & \m4.8$\sigma$ \\
HERMES \cite{hermes} &$\gamma^* d  \rightarrow \Theta^+ X$ &  \m1528 $\pm 2.6 \pm 2.1$ & $17\pm9$ & \m4-6$\sigma$ \\
ITEP \cite{neutrino} &$\nu A  \rightarrow \Theta^+ X$ & \m1533 $\pm 5$ & $<20$ & \m6.7$\sigma$ \\
SVD-2 \cite{svd} &$p A  \rightarrow \Theta^+ X$ & \m1526 $\pm 3$ & $<24$ & \m5.6$\sigma$ \\
COSY \cite{cosy} &$p p  \rightarrow \Theta^+ \Sigma^+$ & \m1530 $\pm 5$ & $<18$ & \m4-6$\sigma$ \\
ZEUS \cite{zeus} &$\gamma^* p  \rightarrow \Theta^+ X$ &  \m1522 $\pm 3$ & $8\pm4$ & \m4.6$\sigma$ \\
\hline
\end{tabular}\\
\end{table}

\begin{figure}[ht]
\begin{center}
\includegraphics[width=8.5cm]{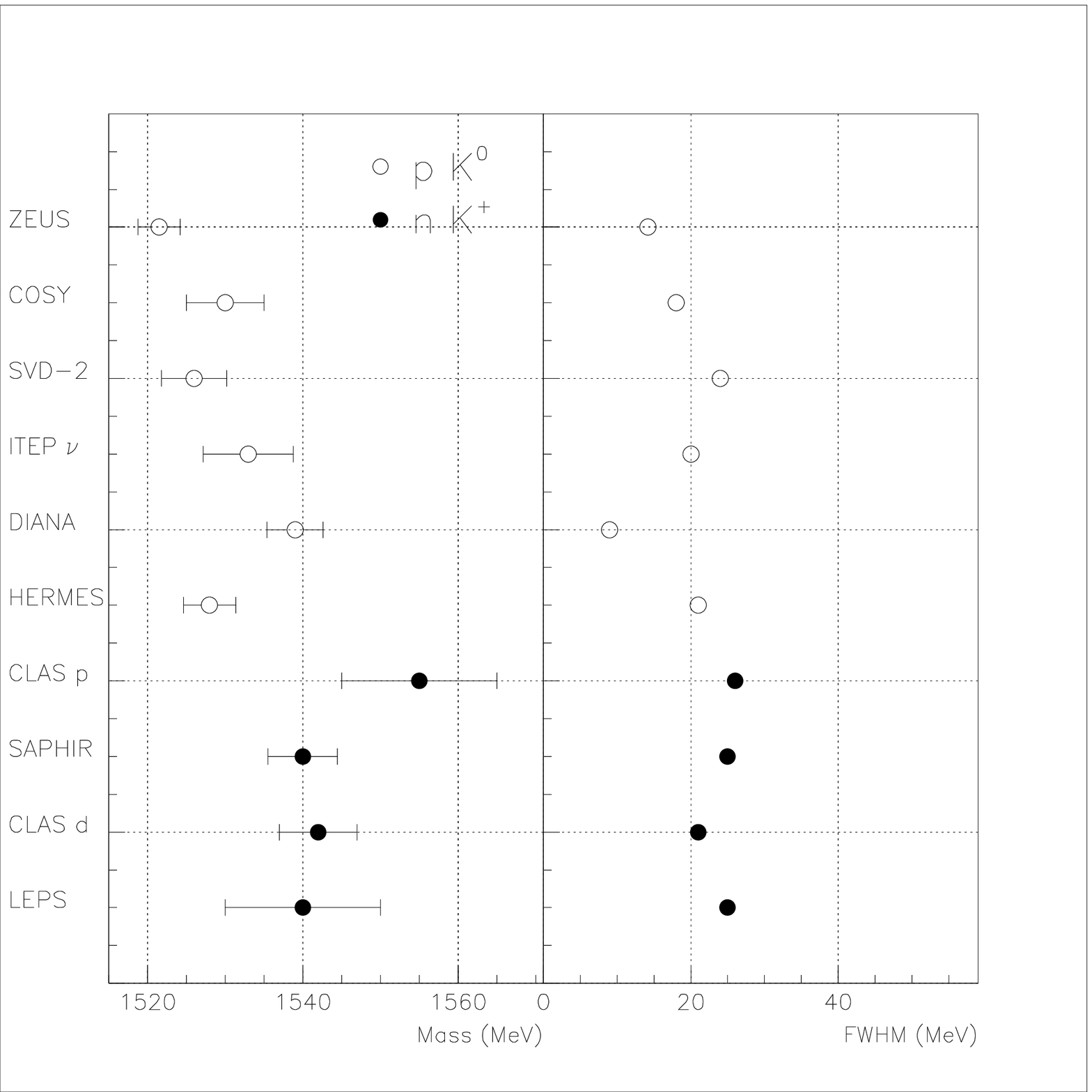}
\caption{Mass and full width of the observed $\Theta^+$ signals.}
\label{fig:teta}
\end{center}
\end{figure}

As already mentioned, the first evidence for the existence of $\Theta^+$ was reported by the LEPS experiment at Spring-8 \cite{leps}.
Studying the reaction $\gamma ^{12}C \rightarrow K^+ K^- X$, they found the 4.6$\sigma$ peak in the $n K^+$ invariant mass showed in fig.~\ref{fig:leps}.
Cuts were applied to select quasi-free processes on the neutron. After Fermi momentum correction, the mass of this peak was determined to be ($1540 \pm 10$) MeV.
The observed width was compatible with experimental resolution, then an upper limit of 25 MeV was derived. This result was closely followed by the DIANA collaboration \cite{diana} who re-analyzed old xenon bubble-chamber data and found a narrow peak in the $K^0 p$ effective mass spectrum in the charge-exchange reaction $K^+ Xe \rightarrow K^0 p Xe^{'}$. Soon thereafter the Jefferson Lab (JLab) CLAS collaboration reported a positive result on a deuterium target \cite{ks,stepan}. In a short period $\Theta^+$ signals were found using various projectiles and targets: in photoproduction on proton by the CLAS collaboration \cite{ks,kouba} and the SAPHIR experiment \cite{saphir}; in quasi real photon deuteron and proton scattering by the HERMES \cite{hermes} and ZEUS collaboration \cite{zeus} respectively; in neutrino and anti-neutrino collisions with nuclei from bubble chamber experiments at CERN and Fermilab \cite{neutrino} and in hadron-hadron collision on various targets \cite{svd,cosy}. Fig.~\ref{fig:teta} summarizes the published mass values and width of these experiments. As it can be seen, reported masses in some cases vary by more than the uncertainties given for the individual experiments, ranging from 1522 to 1555 MeV, with the masses obtained from processes involving $nK^+$ signals  in the initial or final states giving on average 10-15 MeV higher values than those in the $pK^0$ channel.
The observed width (FWHM) are comprised between $\sim (10-30)$ MeV, but in all cases they are dominated by the experimental resolution.\\
The relatively small statistical significance in every measurement, possibly explained by the fact that all the results come from the analysis of data taken for other purposes, and the discrepancy in mass determination, are not the only problems to face to overcome the reticence to accept the existence of the pentaquark. In fact, a major problem frequently mentioned, is that several experiments at high statistic and high energies reported negative results in searches for the $\Theta^+$ \cite{hep}. However the different kinematical and experimental conditions between the low energy exclusive experiments and the high energy semi-inclusive experiments do not allow a direct comparison so that the null results do not prove that the positive ones are wrong.\\
To find a definite answer to the question of existence or non-existence of the $\Theta^+$ and of the other 5-quark baryons and, if they do exist, to determinate their intrinsic properties (mass, width, spin, parity) and the reaction mechanism for their production, a second generation experimental programs have been undertaken in different laboratories, among these at Jafferson Lab.\\
In the next paragraph, published results from Jlab using the CLAS detector and a description of the ongoing and planned next generation experiments are reported.

\section{EXPERIMENTAL PROGRAM AT JLAB}
The superconducting electron accelerator at Jefferson Lab delivers a low-emittance, high resolution, 100\% duty-cycle electron beam to three different experimental Halls A, B and C, simultaneously. The maximum energy is 5.8 GeV (with 75\% polarization available) and the maximum current is 200 $\mu$A.
\\
Hall B is equipped with a tagged photon facility~\cite{tagging} and the CLAS detector~\cite{mecking}.
The bremsstrahlung photon beam is produced from the electron beam impinging on a gold foil radiator located 20 meters upstream of the target. The photon energy is measured by detecting the scattered electrons with an array of plastic scintillators placed in the focal plane of the tagger spectrometer with an energy resolution of $0.1E_0~\%$. The tagger system is able to tag photons in the energy range \mbox{$(0.20-0.95)E_0$}.\\
The CLAS spectrometer is built around six superconducting coils producing a toroidal magnetic field symmetric about the beam and oriented primarily in the azimuthal direction. The coils naturally separate the detector into six sectors, each functioning as an independent magnetic spectrometer.
Each sector is instrumented with 3 sets of multi-wire drift chambers for track reconstruction and one layer of scintillator counters, covering the angular range from $8^\circ$ to $143^\circ$, for time-of-flight measurements. The forward region (\mbox{$8^\circ \leq \vartheta \leq 45^\circ$}) contains gas-filled threshold Cherenkov counters and lead-scintillator sandwich-type electromagnetic calorimeters for particle identification.
For two CLAS sectors the coverage of the electromagnetic calorimeters is extended up to polar angles of $70^\circ$.\\
Due to its large acceptance, which allows the detection of several particles in the final state, CLAS is devoted to study exclusive processes in a large kinematic range and with good resolution. Since 1988, several experiments using photon beam and different targets have been completed and in the last year some of the existing CLAS data were reanalyzed to study possible evidence for pentaquark production.
The two positive published results from the CLAS collaboration were obtained using deuterium and hydrogen targets.

\subsection{Published result of $\Theta^+$ on a deuterium target}

\begin{figure}[h]
\begin{minipage}[t]{70mm}
\includegraphics[width=7cm]{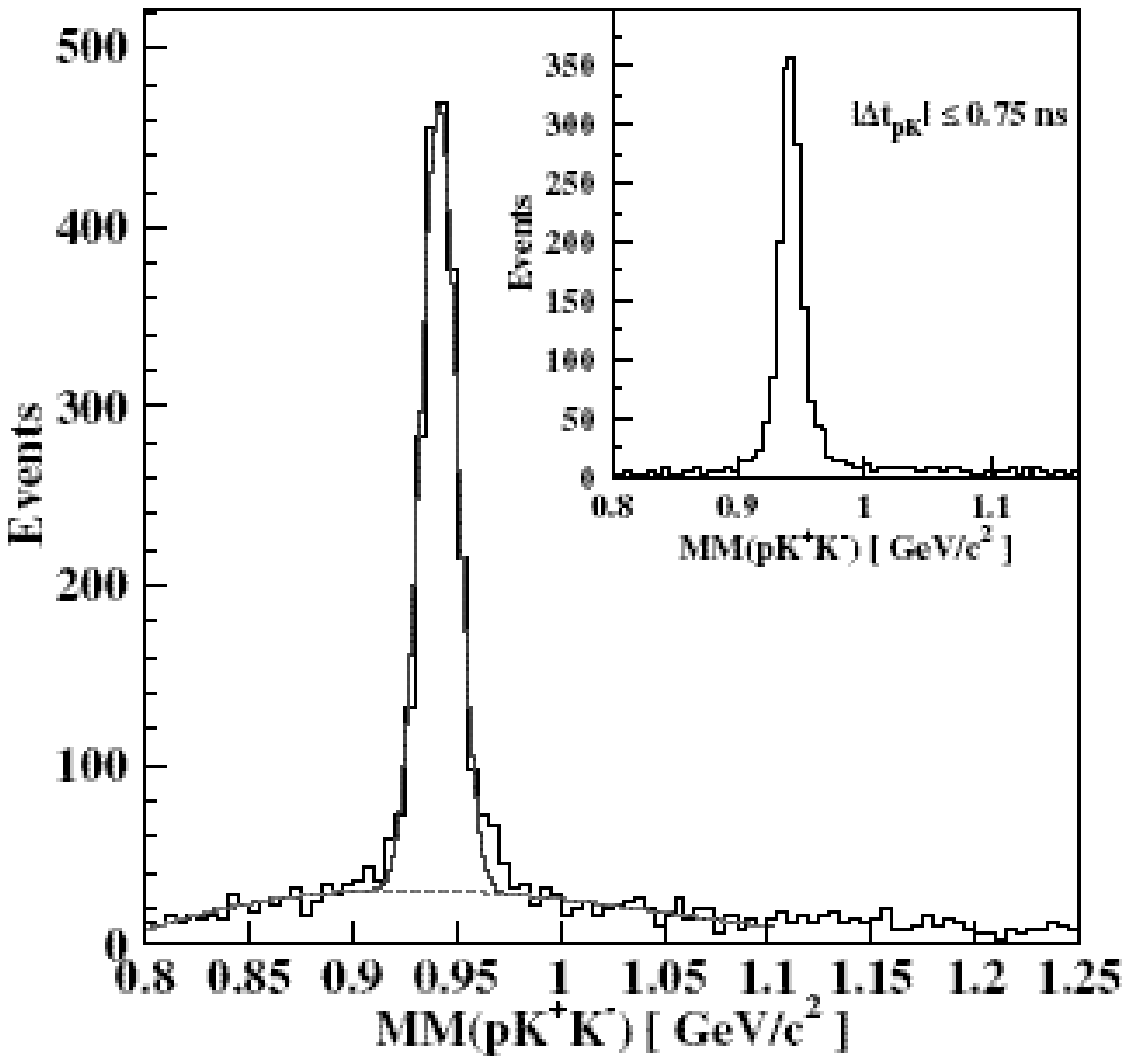}
\caption{Missing mass $M_X$ of $\gamma d  \rightarrow p K^- K^+ X$. A peak at the neutron mass is seen. The inset shows the results with more stringent vertex time cuts.}
\label{fig:graf}
\end{minipage}
\hspace{\fill}
\begin{minipage}[t]{80mm}
\includegraphics[width=7.7cm]{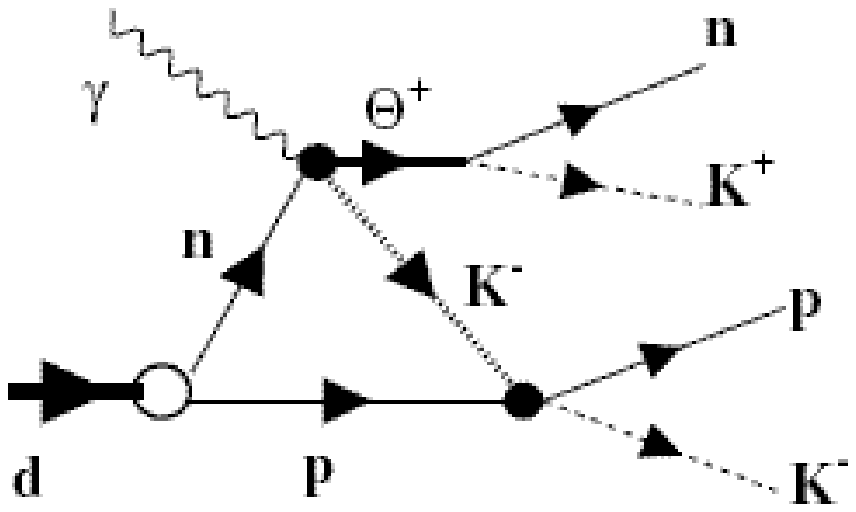}
\caption{Possible diagram for the $\Theta^+$ production from the deuterium target.}
\label{fig:mmn}
\end{minipage}
\end{figure}

\begin{figure}[h]
\begin{minipage}[t]{70mm}
\mbox{}\\
\includegraphics[width=7.2cm]{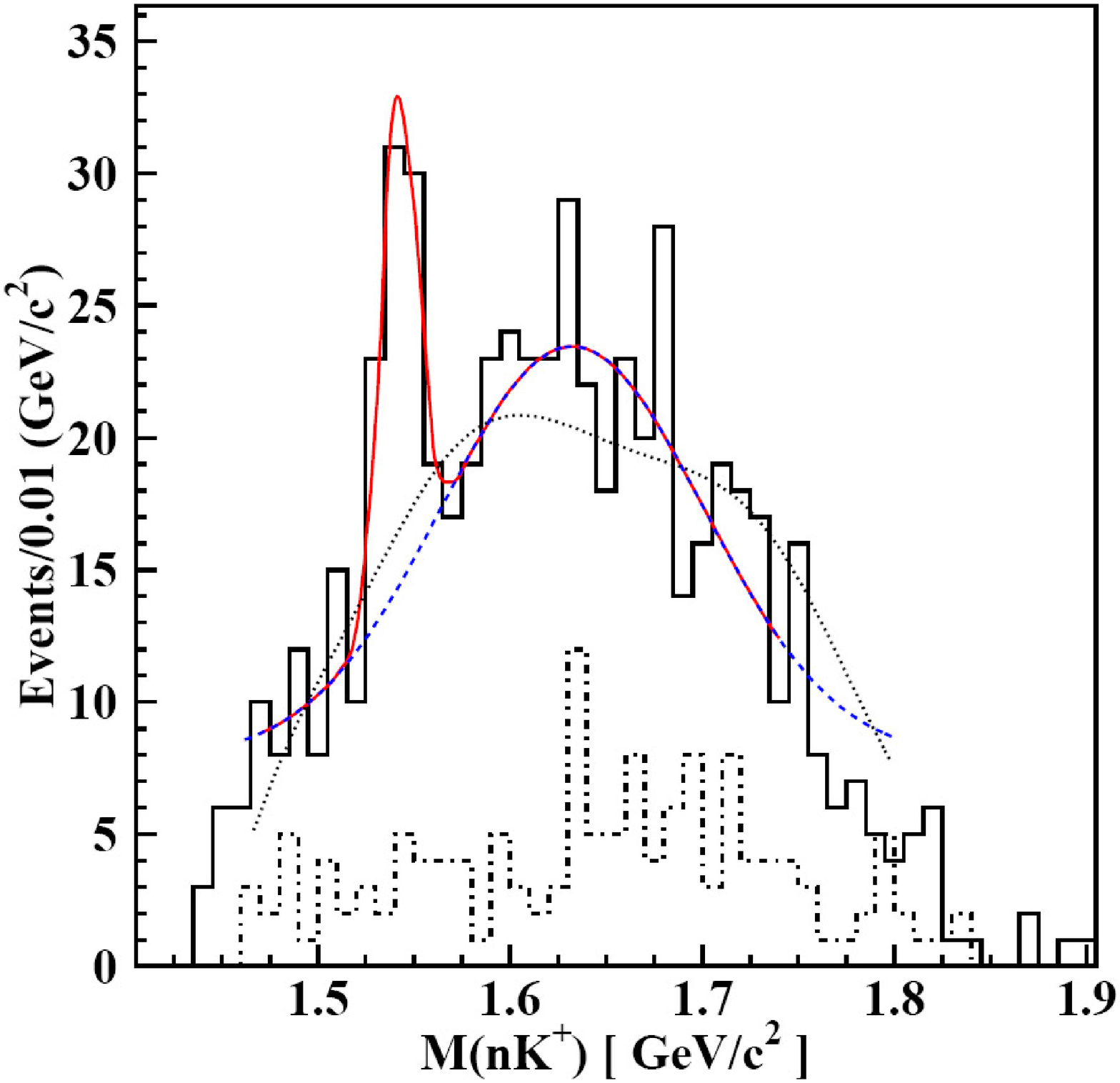}
\caption{$K^+ n$ invariant mass of the CLAS photoproduction experiment on deuteron target (full histogram) \cite{stepan}. Curves are explained in the text.}
\label{fig:thetad}
\end{minipage}
\hspace{\fill}
\begin{minipage}[t]{70mm}
\mbox{}\\
\includegraphics[width=7.2cm]{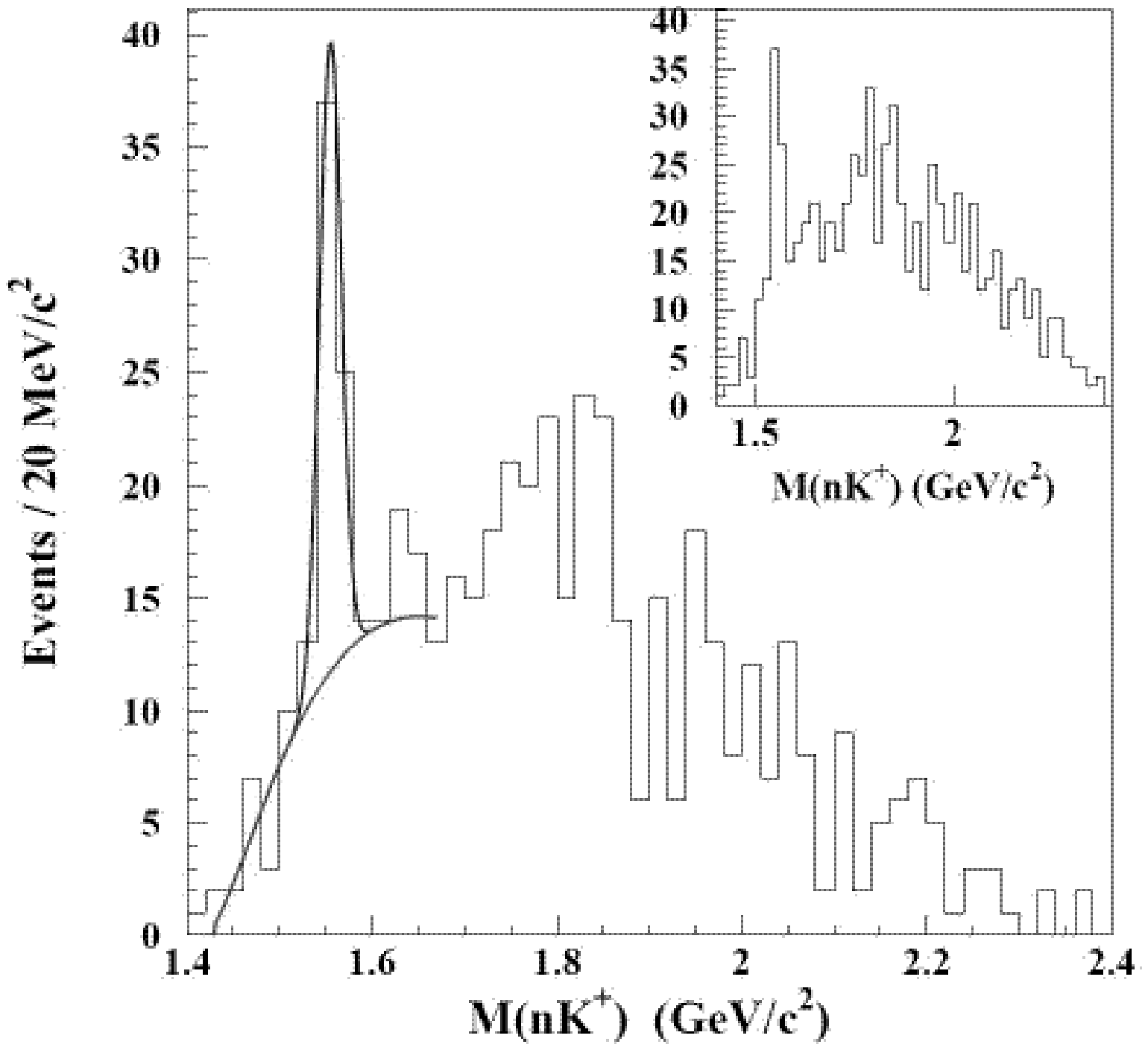}
\vspace{-0.7cm}
\caption{The $M_{nK^+}$ invariant mass spectrum of the CLAS photoproduction experiment on proton target \cite{kouba}.
The inset shows the $nK^+$ invariant mass spectrum with only the $cos \theta^*_{\pi^+} > 0.8$ cut. }
\label{fig:thetap}
\end{minipage}
\end{figure}

Data were taken on August-September 1999 using a 10 cm long deuterium target and tagged photons with maximum energy of 3 GeV. The fully exclusive process  $\gamma d \rightarrow K^- p K^+ n$ was measured. Charged particles 4-momentum vectors were measured with CLAS while the neutron was identified using the missing mass technique after removing the background from pions and protons misidentified as kaons. The missing mass spectrum of $\gamma d \rightarrow K^- p K^+ X $ is shown in fig.~\ref{fig:graf}. To improve the detection of the proton, necessary to insure exclusivity, a final state interaction involving the spectator proton was required to provide it with enough momentum to be detected by CLAS (200 MeV/c is the minimum momentum for CLAS detection). A possible reaction mechanism is illustrated in fig.~\ref{fig:mmn}. Several known processes can contribute to the $pKKn$ events: to eliminate the $\phi$ and $\Lambda(1520)$ cuts in the $K^+K^-$ and $pK^-$ invariant mass were applied. Finally, a cut on the neutron momentum to remove neutron spectator was also applied. The final $K^+n$ invariant mass spectrum, $M(nK^+)$, is shown in fig.~\ref{fig:thetad} along with a fit (solid line) to the peak and a Gaussian plus constant term fit to the background (dashed line). For the fit given, there are 43 counts in the peak at a mass of $1.542\pm0.005~GeV/c^2$ with a width (FWHM) of 0.021 $GeV/c^2$. The statistical significance of the peak is estimated to be 5.2 $\sigma$. The background under the peak was also computed using a Monte Carlo simulation which takes into account the contribution coming from three- ($p K^+ K^-$) and four-body ($p K^+ K^- n$) phase space kaon photoproduction on deuteron (dotted line in fig.~\ref{fig:thetad}).  The statistical significance of the peak using this simulated background is 4.8 $\sigma$. Nevertheless, an analysis of this data using a different technique finds that the significance may not be as large as presented in the published work.

\subsection{Published result of $\Theta^+$ on the hydrogen target}
Data were taken in 1999 and 2000 runs using a 18 cm long hydrogen target and tagged photons with energy in the range $3.0 <E_{\gamma}<5.45$ GeV. The reaction studied was $\gamma p \rightarrow \pi^+ K^- K^+ n$. Also in this case the charged particles were measured with CLAS while the neutron was identified using the missing mass technique. Both the $\gamma p \rightarrow K_0^*\Theta^+$ and $\gamma p \rightarrow \pi^+ K^- \Theta^+$ reaction mechanism were considered, the first with a forward cut in the $K_0^*$ direction and the second with a forward cut in the $\pi^+$ direction. It was found that the signal was most evident in this last case. So, to isolate the  $\pi^+$ t-channel process a cut $cos \theta^*_{\pi^+} > 0.8$ was applied, while to reduce the $K^-$ from t-channel processes all events with  $cos \theta^*_{K^+} < 0.6$ were selected ($\theta^*_{\pi^+}$ and $\theta^*_{K^+}$ are the angles between the $\pi^+$ and $K^+$ mesons and photon beam in the center-of-mass system). The resulting mass spectrum is shown in fig.~\ref{fig:thetap}. The mass of the observed peak is $1555\pm10$ MeV. The statistical accuracy is quoted at about $(7.8 \pm 1.0) \sigma$. 

\subsection{New dedicated experimets in Hall B}
To find a definite answer to the question of the existence of the exotic baryons and, in case of positive signals, to understand their reaction mechanisms and fix their intrinsic properties, a broad experimental programs has been approved in the Hall B. The new dedicated experiments, whose main features are summarized in Table~\ref{table:2}, aim to improve the statistical accuracy of the measurements by at least one order of magnitude.
\begin{table}[htb]
\caption{Summary table of the new dedicated experiments in Hall B at JLab for the search of exotics baryons. }
\label{table:2}
\newcommand{\m}{\hphantom{$-$}}
\newcommand{\cc}[1]{\multicolumn{1}{c}{#1}}
\renewcommand{\tabcolsep}{0.9pc} 
\renewcommand{\arraystretch}{1.2} 
\begin{tabular}{@{}ccccc}
\hline
EXP. &  SEARCH & TARGET & RUNNING & E  (GeV) \\
     &         &        & PERIOD  &                   \\
\hline
E03-113 \cite{stepan2}  & $\Theta^+ $                          & $LD_2$ & 3/13 - 5/16/04   & $E_{\gamma} = 0.8-3.6$\\
(\it g10)           &                                      &   &                  &                       \\
E04-021  \cite{batta}   & $\Theta^+,\Theta{^+}^*,\Theta^{++}$  & $LH_2$ & 5/22 - 7/25/04   & $E_{\gamma} = 0.8-3.8$\\
(\it g11)           &                                      &   &                  &                       \\
E04-010 \cite{elton}    & $\Xi_5$                           & $LD_2$ & 11/20/04 - 1/31/05 & $E_{e^-} = 5.7$     \\
(\it eg3)           &                                      &   &                  &                       \\
E04-017 \cite{dennis}   & $\Theta^+,\Xi_5$                  & $LH_2$ & $2^{nd}$ half 2005 & $E_{\gamma} = 1.5-5.4$\\
(\it super g)           &                                      &   &                    &                      \\
\hline
\end{tabular}\\[2pt]
\end{table}

\subsubsection{Photoproduction of $\Theta^+$ on a deuterium target: {\it g10} experiment}
\vspace{0.3cm}
This experiment run during spring 2004 using a 24 cm length liquid deuterium target, tagged photons in the energy range (0.8 - 3.59) GeV and with the CLAS torus magnetic field set at 2 different values.
The run with lower magnetic field has increased the acceptance for forward ongoing negative particles (which allow us to perform an analysis similar to LEPS for inclusive reactions) while the one with higher magnetic field has the same geometrical acceptance and single track resolution as the published CLAS result on deuterium.
Under this conditions an integrated luminosity of 50 $pb^{-1}$ was achieved (roughly shared between the two magnetic field configurations). This value is a factor 20 greater than the published data, thanks to a longer target, an improved trigger scheme and a longer data taking.
The first pass of the detector calibration and data quality check has already been completed. In fig~\ref{fig:g10} the invariant mass spectra of the $pK^-$ system based on a small fraction of the statistics, give an indication of the quality of the obtained data.
The reaction channels which are object of the analysis are reported in Table~\ref{table:3}. Both decay modes, $pK^0$ and $nK^+$ are being analyzed in missing mass and invariant mass distributions.
\begin{table}[htb]
\caption{Reaction channels under study in the {\it g10} experiment.}
\label{table:3}
\newcommand{\m}{\hphantom{$-$}}
\newcommand{\cc}[1]{\multicolumn{1}{c}{#1}}
\renewcommand{\tabcolsep}{2pc} 
\renewcommand{\arraystretch}{1.2} 
\begin{tabular}{@{}lll}
\hline
REACTION  &  $\Theta^+$ DECAY MODE &  DETECTED           \\
CHANNEL   &                        &  FINAL STATE         \\
\hline
$\gamma d \rightarrow p K^+ K^- n$  & $n K^+$            & $p K^+ K^-$  \\
$\gamma d \rightarrow \Lambda K^+ n$  & $n K^+$            & $p \pi^-K^+ $  \\
$\gamma d \rightarrow \Lambda K^0 p$  & $p K^0_s$            & $p p \pi^-\pi^- \pi^+$ (or only 4, 3, 2 \\
                                     &                      & charged particles)   \\
$\gamma d \rightarrow p p K^0_s K^-$  & $p K^0_s$            & $p \pi^+ \pi^- K^-$  \\
$\gamma ``n'' \rightarrow  K^+ K^- n$  & $n K^+$            & $ K^+ K^-$  \\
\hline
\end{tabular}\\[2pt]
\end{table}

\subsubsection{Photoproduction of $\Theta^+$ on a proton target: {\it g11} and {\it super g} experiments}
\begin{figure}[h]
\begin{minipage}[t]{70mm}
\mbox{}\\
\includegraphics[width=7.4cm]{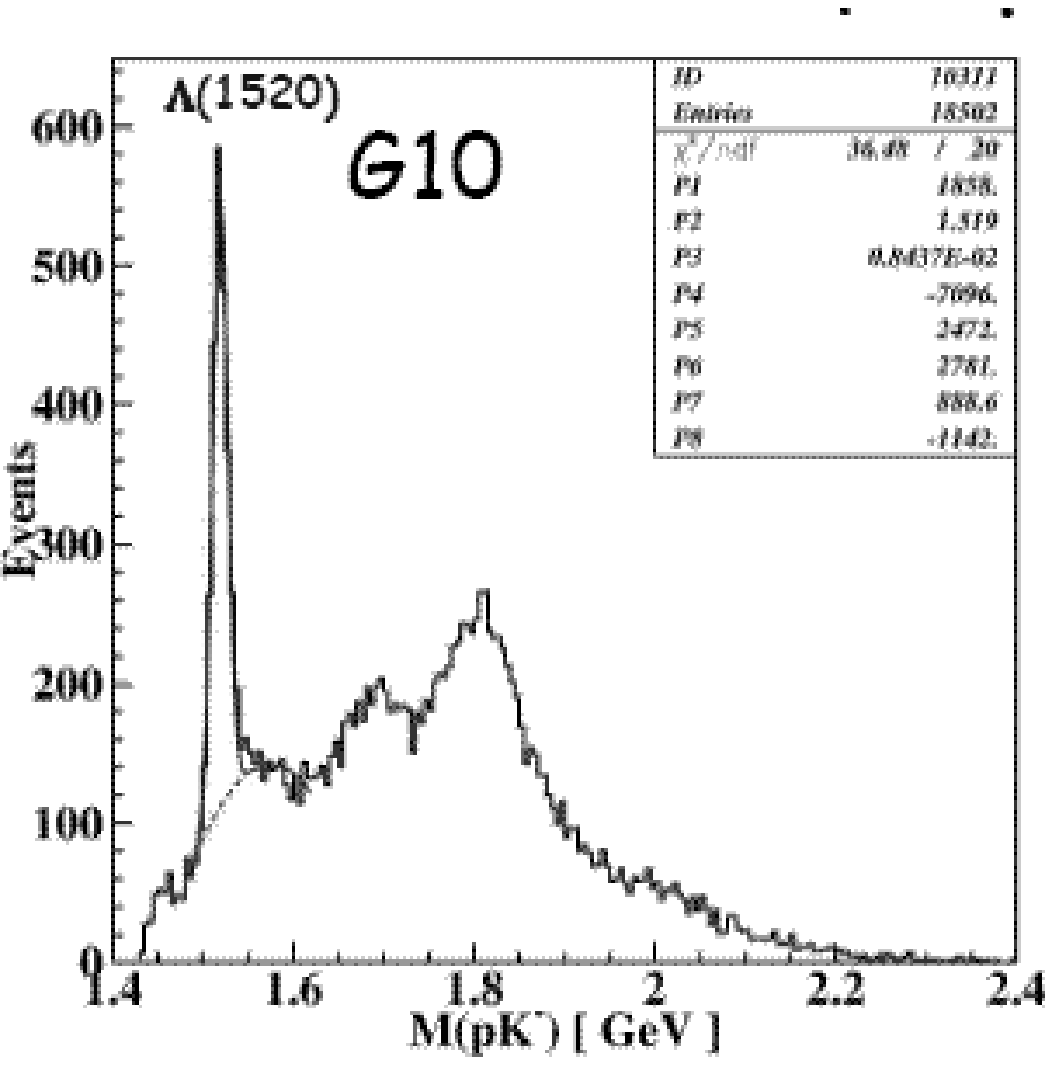}
\vspace{-1cm}
\caption{The $M_{pK^-}$ invariant mass spectrum in the $\gamma d \rightarrow p K^+ K^- n$ reaction, clearly showing the $\Lambda(1520)$ peak. (Preliminary CLAS data.)}
\label{fig:g10}
\end{minipage}
\hspace{\fill}
\begin{minipage}[t]{70mm}
\mbox{}\\
\includegraphics[width=6.9cm]{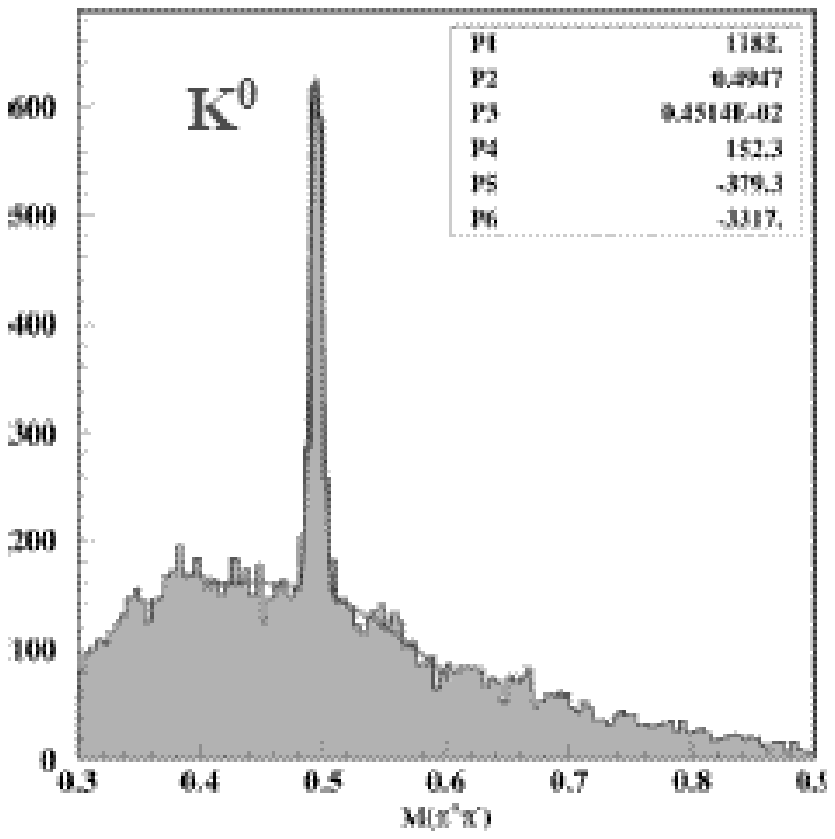}
\vspace{-0.5cm}
\caption{The $M_{\pi^-\pi^+}$ invariant mass spectrum in the $\gamma p \rightarrow \pi^+ \pi^-K^+ n$ reaction, showing the $K^0$ peak. (Preliminary CLAS data.)}
\label{fig:g11}
\end{minipage}
\end{figure}

\vspace{0.3cm}
The g11 experiment run soon after the {\it g10} one and finished to take data at the end of July 2004. Data were taken using a 40 cm length liquid hydrogen target and tagged photons in the enrgy range (0.8 - 3.8) GeV. 
The new longer target, necessary to achieve the goal of this experiment, needed a new start counter detector around the target itself to improve event triggering and particle identification.
Under this conditions an integrated luminosity of 80 $pb^{-1}$ was achieved.
The detector calibration is underway and the data quality check of the CLAS setup is shown in fig.~\ref{fig:g11} where the $M_{\pi^-\pi^+}$ invariant mass spectrum, based on a small fraction of the statistics, in the $\gamma p \rightarrow \pi^+ \pi^-K^+ n$ reaction clearly shows the $K^0$ peak.
The reaction channels under study are: $\gamma p \rightarrow  \bar K^0 K^+ n$, $\gamma p \rightarrow  \bar K^0 K^0 p$, $\gamma p \rightarrow  K^-\pi^+ K^+ n$, $\gamma p \rightarrow K^- \pi^+ K^0 p$, $\gamma p \rightarrow  K^+ \pi^+ \Sigma^-$, $\gamma p \rightarrow  K^+ \pi^- \Sigma^+$, and $\gamma p \rightarrow  K^- K^+ p$.\\
While the goal of the {\it g11} experiment is primarly to check the existence of the $\Theta^+$ and possible excited states on a proton target, the {\it super-g} experiment will be a comprehensive study of exotic baryons from a proton target with a maximum photon energy of about 5.5 GeV. Due to the broad kinematic coverage for a variety of channels, it will measure spin, decay angular distributions and reaction mechanism of the produced particles. Another goal of the {\it super g} experiment is to try to verify the existence of exotic cascades reported by NA49 \cite{alt}. The experiment is scheduled to run in the $2^{nd}$ half of 2005.

\subsubsection{Search for exotic cascades using an untagged virtual photon beam : {\it eg3} experiment}
\vspace{0.8cm}
As mantioned above, observation of other 5-quarks states belonging to the antidecuplet of fig.~\ref{fig:decupletto}, came from NA49 \cite{alt} which found the $\Xi^{--}$ and the $\Xi^0$ at a mass of 1.86 GeV. Nevertheless, up to date, no other experiments have been able to confirm these observations.\\
The goal of the {\it eg3} experiment is to measure the production of pentaquark cascade states using a 5.7 GeV electron beam incident on a thin deuterium target (0.5 cm length) but without detecting the scattered electron. This untagged virtual photon beam is necessary to achieve sufficient sensitivity to the expected small cross sections. In this case, missing mass technique can not be used and the method requires the direct reconstruction of the cascades using their decay products. The sequence of weakly decaying daughter particles provides a powerful tool to pick out the reactions of interest. The main goal of the experiment will be to search for $\Xi^{--}_5 \rightarrow \pi^- \Xi^-$, $\Xi^{-}_5 \rightarrow \pi^0 \Xi^-$ and $\Xi^{-}_5 \rightarrow \pi^- \Xi^0$. Other decay mode are detectable with lower sensitivity. Using the available theoretical estimate for the production cross section of 10 nb, the detection of 460 $\Xi^{--}_5$ particles is expected during a 20 day run. Together with the estimation for the background levels, this represents a statistical significant result of $\sim 20 \sigma$. The experiment is schedule to start to take data in November 2004.

\section{CONCLUSION}
A key question in non-perturbative QCD is the structure of hadrons. The existence of baryon states beyond the minimal $qqq$ configuration is one of the open questions of strong interaction physics. While such states are not prohibited by QCD, no experimental evidence had been found until recently. The first evidence of a narrow resonance with a quark content ($u u d d  \bar{s}$) and so with strangeness S=+1, named $\Theta^+$, was reported by the LEPS collaboration. This observation has been confirmed by another nine experimental groups, with various projectiles and targets. There are however other experiments where $\Theta^+$ has not been seen. In addition, all the signals have rather low statistical precision and there are inconsistencies in the measured masses and widths. Thus, at this time the existence of a narrow pentaquark state is not fully confirmed.\\
The question of whether pentaquarks exist can only be solved by a second generation high statistics experiments. The CLAS collaboration at Jafferson Lab is currently pursuing this goal. High statistics searches for exotic baryons on hydrogen and deuterium target and in various final states have been started in March 2004 and will going on till next year. Data for two of these experiments are already in hand and results are expected by the end of the year.

\end{document}